\documentclass[aps,prl,reprint,groupedaddress,showpacs]{revtex4-1}
\usepackage{amsmath,amssymb}
\usepackage{graphicx}
\begin{document}
\title{Interplay between Plasmon Luminescence and
       Vibrationally Resolved Molecular Luminescence
       Induced by Scanning Tunneling Microscopy}
\author{Kuniyuki Miwa}
\author{Mamoru Sakaue}
\author{Hideaki Kasai}
\email{kasai@dyn.ap.eng.osaka-u.ac.jp}
\affiliation{Department of Applied Physics,
Osaka University,
2-1 Yamadaoka, Suita, Osaka 565-0871, Japan}
\date{\today}
\begin{abstract}
Effects of coupling between a molecular exciton and a surface plasmon (exciton-plasmon coupling) on the luminescence properties of the molecule and the surface plasmons are investigated using the nonequilibrium Green's function method.
Molecular absorption and enhancement by molecular electronic and vibrational modes (molecular modes) lead to dip and peak structures in the luminescence spectra of the surface plasmons.
These structures will correspond to the peak and dip structures observed in a recent experiment.
We found that in addition to the molecular dynamics, the re-absorption by the surface plasmons plays important roles in determining the luminescence spectral profiles.
\end{abstract}
\maketitle
Fundamental understanding of the role of surface plasmons
in the molecular luminescence process
is currently a topic of great interest
owing to its wide applicability in
nanoscale spectroscopy, biosensors, light-emitting devices, etc
~\cite{Tam2007,Moskovits1985,Nie1997,Pettinger2004,
Lakowicz2001a,Fujiki2010a}.
One of the most promising approaches for deepening our knowledge
is to investigate light emission induced by tunneling current of
a scanning tunneling microscope (STM).
STM-induced light emission (STM-LE) spectroscopy has a unique advantage
of having the potential to unveil optical properties
with submolecular spatial resolution.
Moreover, it can remove any complexity
resulting from disturbance of the incident light.
\par
STM-LE from a metal surface was first reported
by the group of Gimzewski in 1988~\cite{Gimzewski1988}. Now, it is well confirmed that
light emission from clean metal surfaces is due to
the radiative decay of surface plasmons
localized near the tip-sample gap region
~\cite{Berndt1991a,Uehara1999,Hoffmann2001,Rossel2010}.
When a molecule is inserted into the gap region,
there are two radiative processes,
i.e., luminescence from the surface plasmons
and the molecule.
If the molecule is directly adsorbed
on a metal substrate,
charge and energy transfer between them
lead to the quenching of molecular luminescence.
Thus, the molecule acts as a spacer
between the tip and the metal substrate,
which slightly modifies the plasmon-mediated emission
~\cite{Hoffmann2002}.
If the molecule is electrically decoupled
from the metal substrate by dielectric films
or molecular multilayers,
intrinsic molecular luminescence,
which is associated with
the electronic and vibrational transitions in the molecule,
can be observed
~\cite{Qiu2003,Dong2004,Ino2008,Chen2010}.
\par
Since close proximity of the tip to the metal substrate induces an intense electromagnetic field generated by the surface plasmons, effects of the interaction between the intense electromagnetic field and the transition moments for the molecular excitations and de-excitations are expected to occur. Due to the fact that the molecular luminescence can be in the same energy window as the plasmon-mediated emission, light emission can be induced by the two kinds of radiative processes simultaneously. Recent experiments have shown that the plasmon-mediated emission overlapping with the molecular absorption enhances molecular luminescence intensity~\cite{Uemura2007,Liu2007a,Liu2009}.
Selective enhancement of luminescence intensity by spectrally tuning the surface plasmon mode to match with a particular transition in the molecule has been observed~\cite{Dong2009a}. Moreover, fascinating new phenomena, i.e., luminescence associated with radiative decay from higher vibrational levels for the excited electronic state of the molecule (hot luminescence) and luminescence at energies beyond the applied bias voltage (upconverted luminescence) have also been observed~\cite{Dong2009a}.
\par
There are two dominant mechanisms involved in the molecular excitations in STM-LE from molecules on metal substrates. One is the excitation by the injection of electrons and/or holes from the electrodes. The other is the excitation by the absorption of surface plasmons that are excited by the inelastic tunneling between the tip and the substrate. Recent experiments suggest the importance of the plasmon-mediated excitations ~\cite{Uemura2007,Liu2007a,Liu2009,Dong2009a,Zhang2012}. In previous theoretical studies, molecular dynamics have been investigated within the framework of the one-body problem, where the surface plasmon is treated as a classical electric field~\cite{Xu2004,Tian2011a,Tian2011}. Tian \textit{et al.}~\cite{Tian2011a,Tian2011} used the plasmon-mediated excitation mechanism to explain some spectral features observed in Ref. ~\onlinecite{Dong2009a}. The results indicate that the molecular dynamics induced by the surface plasmons play essential roles.\par
Direct experimental evidences of the plasmon-mediated excitations of the molecules have been obtained in luminescence spectra acquired with molecule-covered tips over clean metal surface~\cite{Schneider2012}. Although no electron tunneling to the molecules takes place, the observed spectra that can be considered as the luminescence spectra of the surface plasmons show peak and dip structures, some of which match peaks in molecular luminescence and absorption spectra. Thus, the molecular dynamics including the molecular luminescence and absorption have an impact on luminescence-spectral profiles of the surface plasmons. To understand this from a microscopic point of view, there is a need to investigate the interplay between the dynamics of the molecule and the surface plasmons within the framework of the quantum many-body theory.\par
In this study, we investigate the effects of coupling between a molecular exciton and the surface plasmon (exciton-plasmon coupling) with the aid of the nonequilibrium Green's function method~\cite{Keldysh1965} as well as coupling between electronic and vibrational degrees of freedom on the molecule (exciton-vibron coupling). Our results show that the enhancement and suppression of the luminescence intensities of the molecule are due to the enhancement by the surface plasmons and due to the re-absorption of the surface plasmons by the molecule, respectively. Absorption and enhancement by molecular electronic and vibrational modes lead to dip and peak structures in the luminescence spectra of the surface plasmons. In addition to the molecular dynamics, it was found that the re-absorption by the surface plasmons leads to a dent structure in the luminescence spectra of the surface plasmons. The results also show evidence that vibrational excitations assist the occurrence of the upconverted luminescence.
\par
\begin{figure}
\includegraphics[width=3.3cm,clip]{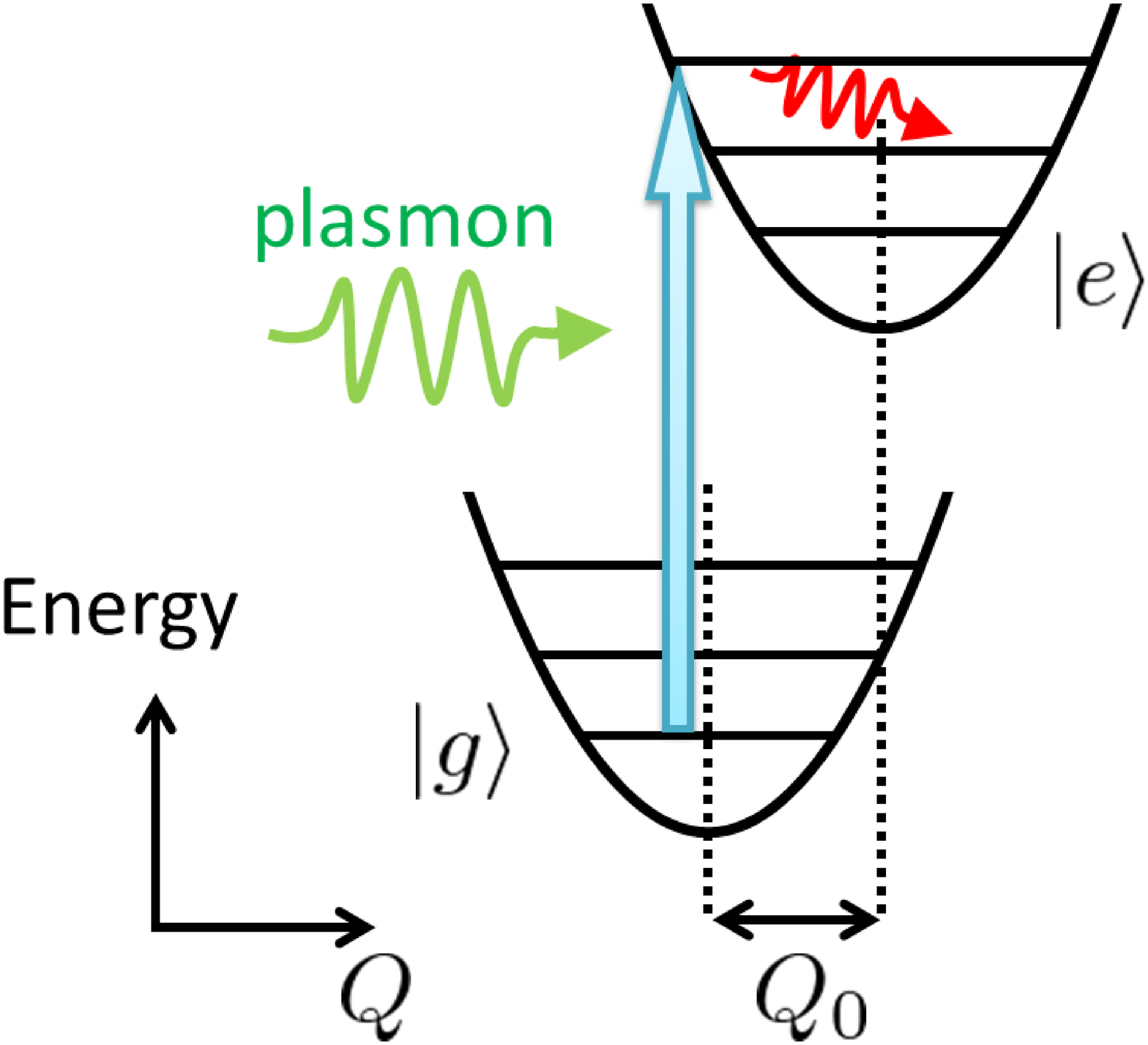}%
\caption{\label{fig:1}
(Color online) Schematic energy diagram of excitation in the absorbed molecular layer whose excitation energy is close to the surface plasmon mode. Horizontal lines in each parabola denote vibrational sublevels where $|g\rangle$ and $|e\rangle$ denote the ground and first-excited electronic states, respectively. The variable $Q$ denotes the vibrational coordinate whose equilibrium position is displaced by $Q_0$ upon the creation of the molecular exciton.}
\end{figure}

A schematic illustration of our model is shown in Fig. \ref{fig:1}.
The electronic degrees of freedom of the molecule are modeled
as a system with
the ground electronic state $|g\rangle$
and the first-excited electronic state $|e\rangle$.
The molecular exciton
interacts with the molecular vibrations and with the surface plasmons.
For simplicity, the molecular vibrations are approximated by a single harmonic oscillator~\cite{Tian2011}.
The surface plasmon modes localized near the tip-sample gap region are approximated by a single energy mode~\cite{Dong2009a,Tian2011,Pettinger2007}.
The creation (annihilation) of the molecular exciton is induced by the absorption (emission) of the surface plasmon~\cite{Schneider2012}.
The Hamiltonian of the system is
\begin{eqnarray}
	H  &=&  \epsilon_{ex} d^\dagger d
	   		+ \hbar \omega_0 b^\dagger b
	   		+ \hbar \omega_p a^\dagger a
	   		+ \sum_{\beta} \hbar \omega_\beta b^\dagger_\beta b_\beta
	\nonumber
	\\
	& &     + M Q_b d^\dagger d
	   		+ V \left( a d^\dagger
	   		+          \mathrm{H.c.} \right)
	   		+ \sum_{\beta} U_\beta Q_b Q_\beta ,
\end{eqnarray}
{\noindent}where $d^\dagger$ and $d$ are creation and annihilation operators for the molecular exciton with energy $\epsilon_{ex}$.
Operators $b^\dagger$ and $b$ are creation and annihilation operators for a molecular vibrational mode with energy $ \hbar \omega_0 $,
$a^\dagger$ and $a$ are for a surface plasmon mode with energy $ \hbar \omega_p $, $b^\dagger_\beta$ and $b_\beta$ are for a phonon mode in the thermal phonon bath, $Q_b=b+b^\dagger$ and $Q_\beta=b_\beta+b^\dagger_\beta$.
The energy parameters $M$, $V$ and $U_\beta$ correspond to the exciton-vibron coupling, the exciton-plasmon coupling, and the coupling between the molecular vibrational mode and a phonon mode in the thermal phonon bath.
\par
By applying the canonical (Lang-Firsov) transformation~\cite{Galperin2006b},
$H$ becomes
\begin{eqnarray}
	\Tilde{H}
		=& & \Tilde{\epsilon}_{ex} d^\dagger d
	   		+ \hbar \omega_0 b^\dagger b
	   		+ \hbar \omega_p a^\dagger a
	   		+ \sum_{\beta} \hbar \omega_\beta b^\dagger_\beta b_\beta
	   		\nonumber \\
	   	 & & + V \left(  a  d^\dagger       X^\dagger
	   		          + \mathrm{H.c.} \right)
	   		+ \sum_{\beta} U_\beta Q_b Q_\beta ,
	   		\label{LFtransform}
\end{eqnarray}
where
$ X  = \exp \left[ - \lambda \left( b^\dagger - b \right) \right]$ ,
$\Tilde{\epsilon}_{ex} = \epsilon_{ex} - M^2/(\hbar \omega_0)$ and
$\lambda = M / (\hbar \omega_0)$. \par
The luminescence spectra of the molecule and the surface plasmons are obtained by the relations, $B_L(\omega)=-\Im L^< (\omega) / \pi$ and $B_P(\omega)=-\Im P^< (\omega) / \pi$~\cite{Galperin2009}. Here $L^<$ and $P^<$ are the lesser projections of the Green's function of the molecular exciton and the surface plasmon on the Keldysh contour ($L$ and $P$), respectively.
\par
To investigate the excitation properties and analyze the structures in the luminescence spectra, the spectral functions of the molecule and the surface plasmons, which correspond to their absorption spectra, are calculated by the relations, $A_L(\omega)=-\Im L^r (\omega) / \pi$ and $A_P(\omega)=-\Im P^r (\omega) / \pi$, where $L^r$ and $P^r$ are the retarded projections of $L$ and $P$.
\par
To obtain the spectral functions and the luminescence spectra, it is needed to calculate $L$ and $P$, which are defined as
\begin{eqnarray}
L(\tau,\tau')
	&=&  \frac{1}{i \hbar}
	\langle T_C \{
	d (\tau) X(\tau) d^\dagger (\tau') X^\dagger(\tau')
	\} \rangle _{\Tilde{H}} ,\\
P(\tau,\tau')
	&=&  \frac{1}{i \hbar}
	\langle T_C \{
	a(\tau) a^\dagger(\tau')
	\} \rangle _{\Tilde{H}} ,
\end{eqnarray}
where $\langle\cdots\rangle _{\Tilde{H}}$ denotes
statistical average in representation
by system evolution for $\Tilde{H}$.
The variable $\tau$ is the Keldysh contour time variable.
The operator $T_C$ denotes the time ordering along the Keldysh contour.
\par
To calculate the Green\rq{}s function $L$ containing the exciton-plasmon coupling $V$, we take into account a random-phase-approximation (RPA)-type diagrammatic series. This approximation is expected to hold for $M \gg V^2$~\cite{Maier2011}. The integral equation for $L$ is given by
\begin{eqnarray}
	L(\tau,\tau')
	& & = L_\mathrm{b} (\tau,\tau') \nonumber \\
	 & & +\int d\tau_1 d\tau_2
	   L_\mathrm{b} (\tau,\tau_1)
	   |V|^2 P^{(0)}(\tau_1, \tau_2) L(\tau_2,\tau'),
	\label{eq:polarization}~~
\end{eqnarray}
where
\begin{eqnarray}
	L_\mathrm{b} (\tau,\tau')
	&=& L_\mathrm{el}(\tau,\tau') K(\tau,\tau'), \\
	L_\mathrm{el} (\tau,\tau') &=& \frac{1}{i\hbar}
	        \langle T_C \{ d (\tau) d^\dagger (\tau') \}\rangle_{\Tilde{H}}, \\
	K(\tau,\tau') &=& \langle T_C \{
		X(\tau) X^\dagger(\tau') \} \rangle_{\Tilde{H}},
\end{eqnarray}
{\noindent}and $P^{(0)}$ is the plasmon Green's function for $V=0$. Here, we employ an approximation of decoupling electronic and vibrational parts in $L_\mathrm{b}$, and calculate $L_\mathrm{el}$ and $K$.
The integral equation for $L_\mathrm{el}$ is given by
\begin{eqnarray}
	L_\mathrm{el}(\tau,\tau')
	&=& L^{(0)}(\tau,\tau') \nonumber \\
	&+& \int d\tau_1 d\tau_2
		  L^{(0)}(\tau,\tau_1)
		  \Sigma (\tau_1, \tau_2)
		  L_\mathrm{el}(\tau_2,\tau'),~ ~
	\label{eq:L_ele}\\
	\Sigma (\tau_1, \tau_2)
	&=& |V|^2 P^{(0)} (\tau_1, \tau_2) K(\tau_2,\tau_1),
\end{eqnarray}
where $L^{(0)}$ is $L_\mathrm{el}$ for $V=0$. The integral equation for $P$ is given by
\begin{eqnarray}
	P(\tau,\tau')
	& & = P^{(0)} (\tau,\tau') \nonumber \\
	 & & +\int d\tau_1 d\tau_2
	   P^{(0)} (\tau,\tau_1)
	   |V|^2 L(\tau_1, \tau_2) P(\tau_2,\tau').
	\label{eq:plasmon}~~~~
\end{eqnarray}
We express the correlation function $K$ in terms of the vibrational Green's function $D$, which is defined by
\begin{equation}
	D(\tau,\tau') = \frac{1}{i\hbar}
	    \langle T_C\{ P_b (\tau) P_b (\tau') \}\rangle_{\Tilde{H}}
\end{equation}
where $P_b=-i\left(b-b^\dagger\right) $
~\cite{Galperin2006b}.
The self-energy for $D$ can be expressed by using $P$ and $L$.\par
By assuming the condition of stationary current, the distribution function $N_\mathrm{pl}$ of the surface plasmons excited by inelastic tunneling between the tip and the substrate is given by
\begin{equation}
	N_\mathrm{pl}(\omega)
	= \begin{cases}
	T_\mathrm{pl} \left(1-\left| \frac{ \hbar \omega }
	                     { eV_\mathrm{bias} } \right| \right),
	& \left| \hbar \omega \right| <
	              \left| eV_\mathrm{bias} \right| \\
    0,	& \text{others}
	  \end{cases}
	,
	\label{N_Pla}
\end{equation}
where $T_\mathrm{pl}$ is a coefficient related to the current amplitude due to the inelastic tunneling ~\cite{Persson1992}, $e$ is the elementary charge, and $V_\mathrm{bias}$ is the bias voltage.
The lesser and greater projections of $P^{(0)}$ are obtained through the relations, $P^{(0),<} (\omega)=2iN_\mathrm{pl}(\omega)\Im{P^{(0),r}(\omega)}$ and $P^{(0),>} (\omega)=2i\left[ 1 + N_\mathrm{pl}(\omega) \right] \Im{P^{(0),r}}$, where $P^{(0),r}$ is the retarded projection of $P^{(0)}$~\cite{Keldysh1965}.
The equations mentioned above are self-consistently solved. After convergence, the spectral functions and the luminescence spectra are calculated.
\par
The parameters given correspond to the experiment on STM-LE from tetraphenylporphyrin(TPP) molecules on the gold surface~\cite{Dong2009a}:
$\Tilde{\epsilon}_{ex} = 1.89~\mathrm{eV}$,
$\hbar \omega_0 = 0.16~\mathrm{eV}$
and $\hbar\omega_p=2.05~\mathrm{eV}$.
The statistical average is taken for
temperature $T=80~\mathrm{K}$~\cite{Dong2009a}.
The square of $\lambda$
has been reported to be 0.61
on the basis of first-principles calculations~\cite{Tian2011}.
The parameter $U_\beta$ is given so that
the molecular vibrational lifetime due to the coupling to
the thermal phonon bath is 13 ps~\cite{Dong2009a}.
A Markovian decay is assumed for the surface plasmon
so that the plasmon lifetime for $V=0~\mathrm{eV}$
becomes 4.7 fs~\cite{Dong2009a,Tian2011}.
The coefficient $T_\mathrm{pl}$ is set to $10^{-4}$
to give the tunneling current $I_t=200~\mathrm{pA}$.
\par
\begin{figure}
\includegraphics[width=8.6cm,height=6.4cm,keepaspectratio]{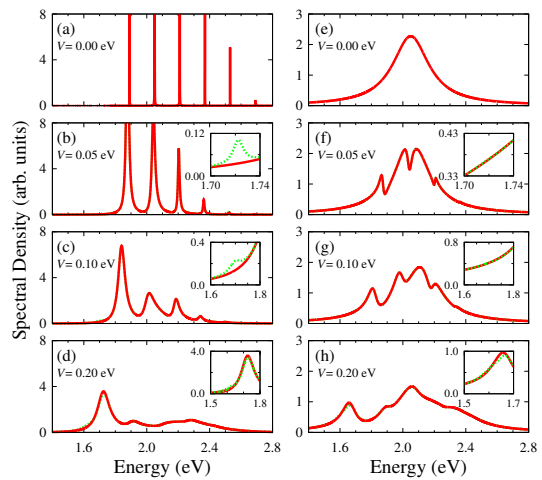}%
\caption{\label{fig:2}
(Color online) Spectral functions of the molecular exciton $A_L$ for (a) $V=0~\mathrm{eV}$, (b) $V=0.05~\mathrm{eV}$, (c) $V=0.10~\mathrm{eV}$ and (d) $V=0.20~\mathrm{eV}$, and the surface plasmon $A_P$ for (e) $V=0~\mathrm{eV}$, (f) $V=0.05~\mathrm{eV}$, (g) $V=0.10~\mathrm{eV}$ and (h) $V=0.20~\mathrm{eV}$. Red solid and green dashed lines show functions for the bias voltage $V_\mathrm{bias}=1.8~\mathrm{V}$ and $2.5~\mathrm{V}$, respectively.}
 \end{figure}

Figure \ref{fig:2} shows the calculated spectral functions $A_L$ and $A_P$.
The peak positions for both $A_L$ and $A_P$ are shifted due to the exciton-plasmon coupling $V$, remarkably for the lowest one upon excluding the small structure near 1.7~eV.
It is confirmed that
the amount of the shift is roughly scaled by $V^2$
in the range from $V=0.05~\mathrm{eV}$ to $V=1~\mathrm{eV}$.
These peaks include the contribution from
the processes of electronic transitions
accompanied by vibrational state transitions.
The amplitude of the asymmetric structure in the vicinity of 1.7~eV
increases as the bias voltage $V_\mathrm{bias}$ increases.
These structures originate from a process which involves
energy transfer from the vibrational state.
\par
\begin{figure}
\includegraphics[width=8.6cm,clip]{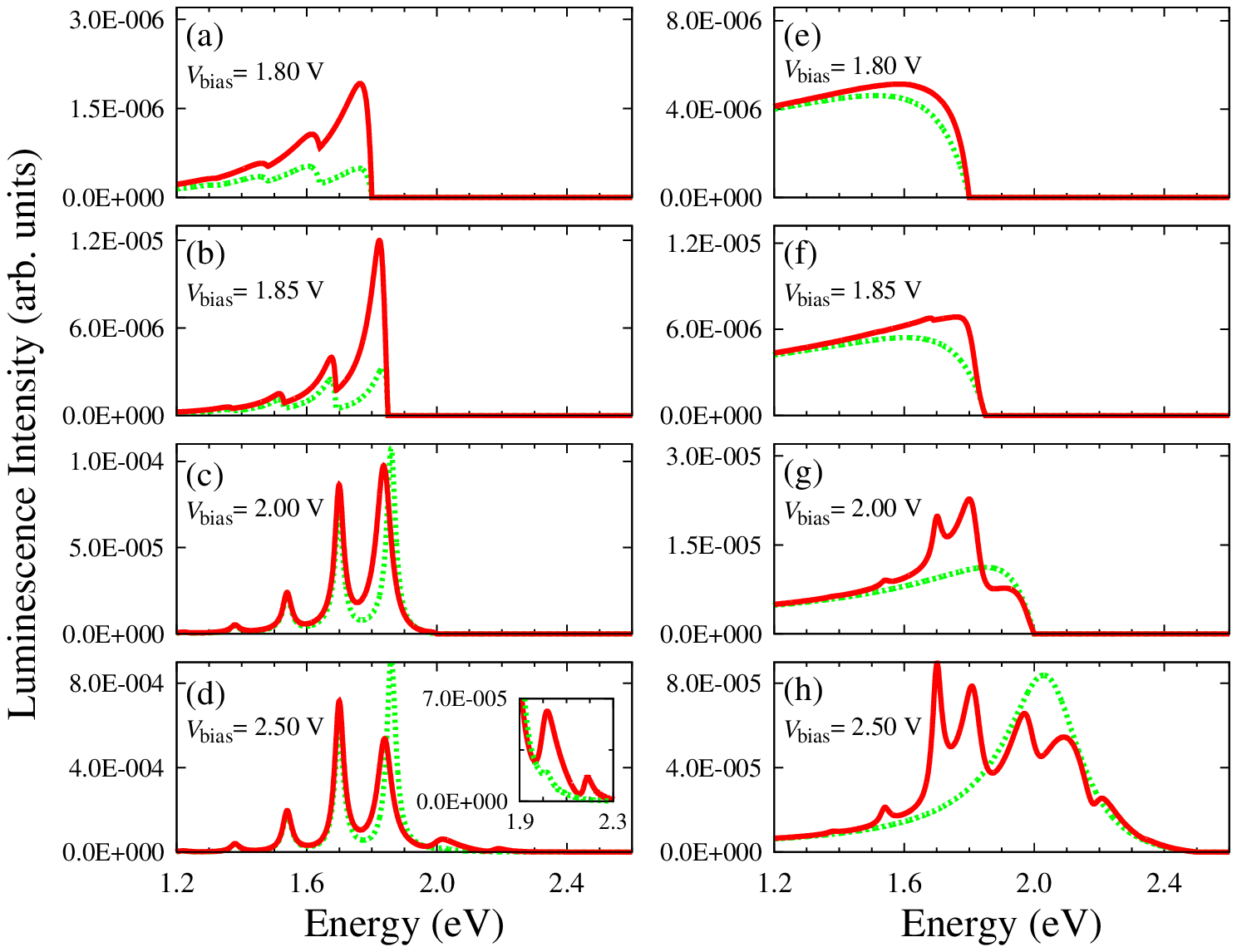}
\caption{\label{fig:3}
(Color online) Luminescence spectra of the molecule $B_L$ for the bias voltage (a) $V_\mathrm{bias} = 1.8~\mathrm{V}$, (b) $V_\mathrm{bias} = 1.85~\mathrm{V}$, (c) $V_\mathrm{bias} = 2.0~\mathrm{V}$ and (d) $V_\mathrm{bias} = 2.5~\mathrm{V}$, and the surface plasmon $B_P$ for the bias voltage (e) $V_\mathrm{bias} = 1.8~\mathrm{V}$, (f) $V_\mathrm{bias} = 1.85~\mathrm{V}$, (g) $V_\mathrm{bias} = 2.0~\mathrm{V}$ and (g) $V_\mathrm{bias} = 2.5~\mathrm{V}$. Red solid and green dashed lines in $B_L$ are luminescence spectra for $L$ and $L_b$, respectively. Red solid and green dashed lines in $B_P$ are luminescence spectra for $P$ and $P^{(0)}$, respectively. The exciton-plasmon coupling is $V=0.10~\mathrm{eV}$.}
\end{figure}

The luminescence spectra $B_L$ and $B_P$ are shown in Fig. \ref{fig:3}.
At $V_\mathrm{bias}=1.8~\mathrm{V}$,
the luminescence intensity of the molecule [Fig. \ref{fig:3}(a)]
is lower than the surface plasmons [Fig. \ref{fig:3}(e)].
While at $V_\mathrm{bias}>1.85~\mathrm{V}$,
the luminescence intensity of the molecule
becomes higher than the surface plasmons.
In addition, at $V_\mathrm{bias}=2.5~\mathrm{V}$,
the luminescence spectra $B_L$ [red solid line in Fig. \ref{fig:3}(d)]
shows peak structures near $2.05~\mathrm{eV}$ and $2.21~\mathrm{eV}$.
The position of these peak structures correspond to the energies of transitions from the higher vibrational levels for the excited electronic state to the ground electronic state of the molecule. Therefore, these peak structures are attributed to the hot luminescence, which have been observed in the experiment~\cite{Dong2009a}.
The luminescence intensity of the peak near $1.89~\mathrm{eV}$ is
suppressed.
In this energy range, the high intensity peak is found in
the spectral function of the molecule [Fig. \ref{fig:2}(c)].
Thus, the suppression of the luminescence intensity is attributed to the
re-absorption of the surface plasmons by the molecule (re-absorption process).
The re-absorption processes are classified into two kinds:
one is accompanied by the vibrational excitations
while the other is not
(inelastic and elastic processes, respectively).
From the spectral function of the molecule,
it is confirmed that for $\lambda^2 < 1$,
the elastic process is dominant over the inelastic processes.\par
In the luminescence spectra of the surface plasmons $B_P$, complicated peak and dip structures appear at $V_\mathrm{bias} > 1.85~\mathrm{V}$ [red solid line in Figs. \ref{fig:3}(g), \ref{fig:3}(h)]. The position of the peak structures in the energy range lower than $1.75~\mathrm{eV}$ (near 1.70, 1.54 and 1.38~eV) correspond to the peaks in the luminescence spectra of the molecule $B_L$ [red solid line in Figs. \ref{fig:3}(c), \ref{fig:3}(d)]. These peak structures are, therefore, due to the enhancement by the molecular electronic and vibrational modes (molecular modes).
The position of the dip structures in the energy range higher than $1.8~\mathrm{eV}$ in $B_P$ (near 1.86, 2.02 and 2.18~eV) correspond to the peaks in the spectral function of the molecule $A_L$. These dip structures are attributed to the molecular absorption. These peak and dip structures will correspond to the peak and dip structures observed in the recent experiment~\cite{Schneider2012}. In addition, our results show that a dent structure appears near the energy of the surface plasmon mode (2.05~eV) in $B_P$ [Fig. \ref{fig:3}(h)], although $A_L$ has its maximum intensities near 1.86~eV. It is found that the re-absorption by the surface plasmons leads to the dent structure in the energy range near 2.05~eV, where the spectral function of the surface plasmons $A_P$ has its maximum intensity.
Thus, we insist that in addition to molecular luminescence and absorption, the re-absorption by the surface plasmons play crucial roles in determining the luminescence spectral profiles of the surface plasmons.We expect that these results can be verified by experiments, for example by a comparison between the luminescence spectra acquired with a molecule-covered tip and with a clean metallic tip on a clean metal surface, which correspond to the red solid and green dashed lines in Fig. \ref{fig:3}(h), respectively~\cite{Schneider2012}.
\par
\begin{figure}
\includegraphics[width=8.6cm,clip]{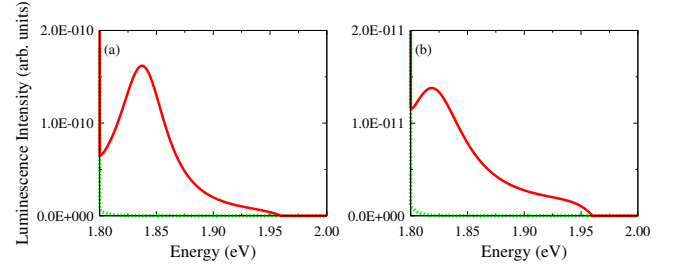}
\caption{\label{fig:4}
(Color online) Luminescence spectra of (a) the molecule $B_L$ and (b) the surface plasmon $B_P$ at the bias voltage $V_\mathrm{bias} = 1.8~\mathrm{V}$. Red solid and green dashed lines show spectra for vibrations in the nonequilibrium and thermal equilibrium states, respectively. The exciton-plasmon coupling is $V=0.10~\mathrm{eV}$.}
\end{figure}

The upconverted luminescence can be seen in Fig. \ref{fig:4}.
Its contribution can be confirmed in comparison with the vibrational state in thermal equilibrium [green dashed line in Fig.~\ref{fig:4}], where the molecular vibration is distributed according to the Bose distribution function at $T=80~\mathrm{K}$ and is therefore almost in the ground state.
The vibrational excitations are accompanying the electronic transitions, which occur via the excitation channels resulting from the exciton-plasmon coupling $V$.
The vibrational occupation number $\langle b^\dagger b \rangle_{\Tilde{H}}$ are scaled by $\lambda^2 |V|^2 n_{ex}/\gamma$, where $n_{ex}$ is the population of the molecular exciton, which is given by $n_{ex}=\langle d^\dagger d \rangle_{\Tilde{H}}$, and $\gamma$ represents the vibrational damping for $V=0~\mathrm{eV}$.
Thus the vibrational excitations assist the occurrence of the upconverted luminescence.
\par
In conclusion, the luminescence properties
of the molecule and the surface plasmons
can be strongly influenced by the interplay between
their dynamics resulting from the exciton-plasmon coupling.
The luminescence spectra of the molecule
are modified due to the re-absorption of the surface plasmons by the molecule.
The peak structures in $B_P$ arise due to the enhancement by the molecular modes and the dip structures due to the molecular absorption.
It is found that the re-absorption by the surface plasmons leads to the dent structure in $B_P$. Thus, in addition to the molecular dynamics, the re-absorption by the surface plasmons plays crucial roles in determining the spectral profiles of $B_P$.
Moreover, our calculations have reproduced the hot luminescence and the upconverted luminescence assisted by the molecular vibrations.
\begin{acknowledgments}
This work is supported in part
by MEXT
through the G-COE
program ``Atomically Controlled Fabrication Technology '',
Grant-in-Aid for Scientific Research on Innovative Areas Program
(2203-22104008) and Scientific Research (c) Program (22510107).
It was also supported in part by JST
through ALCA Program
``Development of Novel Metal-Air Secondary Battery Based
on Fast Oxide Ion Conductor Nano Thickness Film''
and Strategic Japanese-Croatian Cooperative Program on Materials Science
``Theoretical modeling and simulations of the structural,
electronic and dynamical properties of surfaces and nanostructures
in materials science research''.
Some of the calculations presented here were performed using
the ISSP Super Computer Center,
University of Tokyo.
The authors are deeply grateful to
Professor Wilson Agerico Di\~{n}o
and Professor Hiroshi Nakanishi in Osaka University for useful discussions.
\end{acknowledgments}
%
%
%
\end{document}